\documentclass[twocolumn]{aa}
\usepackage{graphicx}
\usepackage{ulem}
\usepackage{txfonts}
\def\ux{UX\,Arietis }
  \def\new#1 {{\bf #1 }}
\begin{document}
\title{Discovery of Solar Rieger Periodicities in Another Star}



   \author{M. Massi
          \and
        J. Neidh\"ofer
        \and
        Y. Carpentier
        \and
        E. Ros
          }

   \offprints{M. Massi}

   \institute{Max-Planck-Institut f\"ur Radioastronomie, 
              Auf dem H\"ugel 69, D-53121 Bonn, Germany\\
              \email{[mmassi,jneidhoefer,ros]@mpifr-bonn.mpg.de, yvain2020@yahoo.fr}
             }

   \date{Draft --- \today
}

   \abstract{
The Rieger periods are solar cycles with a time scale of months,
which are present  in both  flaring activity and  sunspot occurrence.
These short-term periodicities, tentatively explained by   equatorially
trapped Rossby-type waves modulating the emergence of magnetic
 flux at the surface,  are considered a peculiar and not yet 
fully understood solar phenomenon.  We chose a stellar system with solar 
characteristics, UX Arietis, and performed a timing analysis
of two 9-year datasets of
radio and optical observations.
The analysis reveals  a 294-day cycle.
When  the two 9-year datasets are folded with this period,
a synchronization of 
the peak of the optical light curve (i.e., the minimum spot coverage) 
with  the minimum  radio flaring activity
is observed.
This close relationship between  two completly independent curves makes
it very likely that the  294-day cycle is real.
We conclude that the process invoked for the Sun of a
periodical emergence of  magnetic flux
 may also be applied to \ux and can explain the
cyclic flaring activity  triggered by
interactions between successive cyclic  emergences of magnetic flux.
}
   \maketitle
 %

\section{Introduction}
\begin{figure}[htp]
\centering
\includegraphics[height=1.2\textwidth, width=0.5\textwidth]{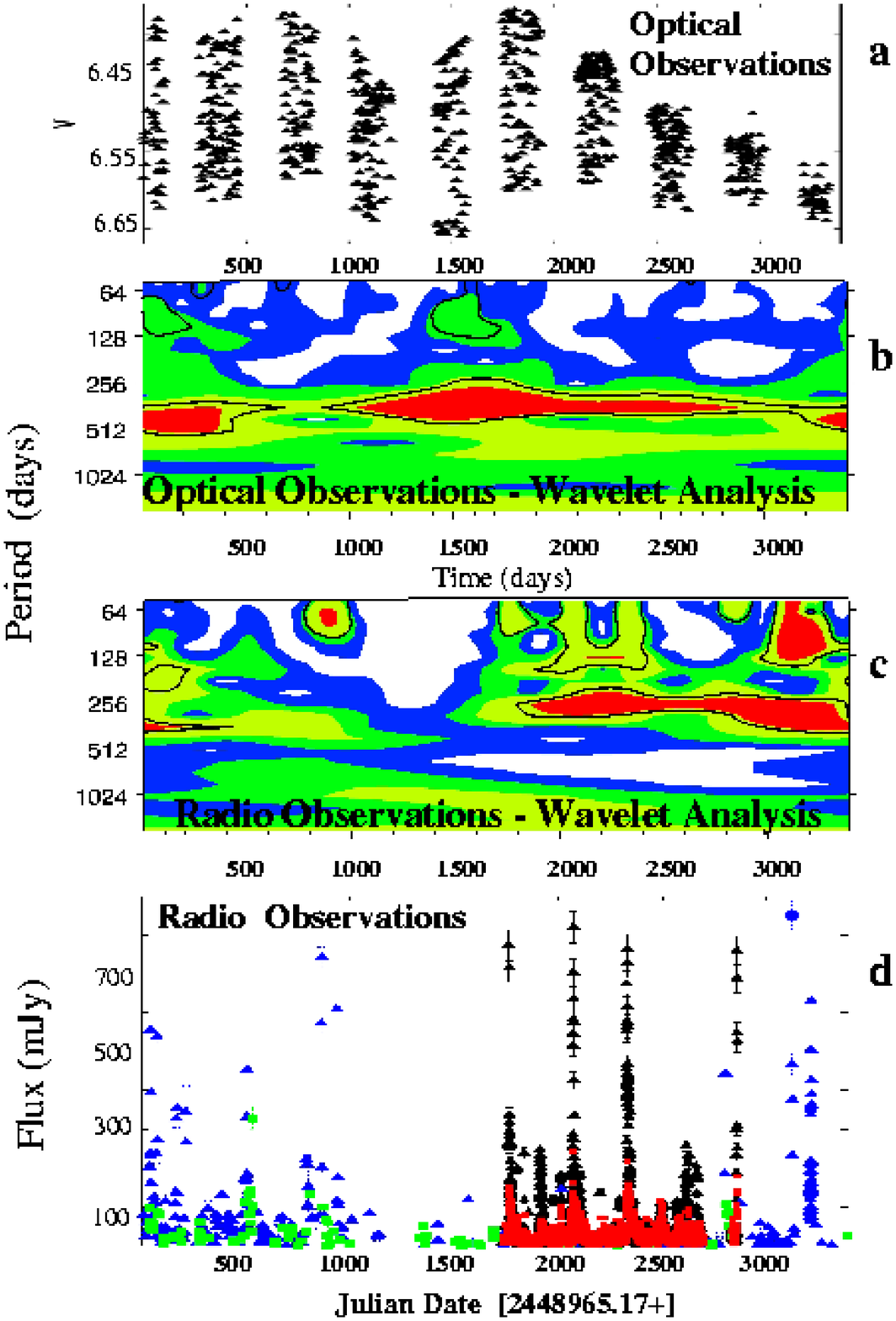}
\caption{
Radio and optical measurements and their corresponding wavelet spectra.
The x-axis of all figures is the same and equal to the  Julian Date [JD$-$2448965.17].
Figures (\new{a}) and (\new{d}) are described in Sect. 2; 
Figures (\new{b}) and (\new{c}) in Sect. 3. 
The  Radio observations (\new{d}) are:
Effelsberg data 
above 8\,GHz (blue triangles; the blue dot is an out of scale  flare of 1318 mJy),  
 and below 8\,GHz (green squares), 
Green Bank Interferometer data
at 8.3\,GHz 
(black triangles), and at 2.25\,GHz (red squares).
}
\label{fig:fig1}
\end{figure}

RS CVn stars are tightly-orbiting binary systems,
consisting of an F or  G type
dwarf and a  G or K type subgiant.
The components of these systems are fast rotators due to the strong  
tidal interaction that synchronizes 
stellar spins and  orbital motion, 
making the dynamo mechanism very effective: 
 Quite  extended (covering up to  several percent of the stellar surface)
cool spots on the  more active star of the system
and intense X-rays and radio flares have been observed
(Hall  1976; Owen et al. 1976; Budding et al. 2002).
Spots and flares are related  to each other:
Solar and stellar spots are created by intense magnetic fields
pushing their way through the photosphere
and as  observed on the Sun, flares are triggered by 
interactions between new and older emergences of magnetic flux  
in the same area (Nishio et al. 1997).

One of the most active RS CVn systems
is  UX~Arietis 
(Beasley \& G\"udel 2000).
The flaring activity of UX Arietis shows some
periodicities:
 3\,-yr monitoring with the Effelsberg 100-m radio telescope 
revealed  a  cycle of several months modulated by shorter
cycles (Massi et al.\ \cite{Mas98}) and 
 later  3\,-yr radio monitoring with the Green Bank
Interferometer not only  confirmed the existence of
such a cyclic flaring activity in UX\,Arietis, but also indicated
periodical  activity in  other stars (Richards et al.\ \cite{Ric03}).

Where do  these periodicities come from ? If
spots and flares are related to each other (like in the Sun), 
does a periodical flaring activity
imply a similar periodicity in the spot occurrence ? 
The best known periodicity in the production of energetic  flares
is the Solar Rieger periodicity of  
152-158 days
(Rieger et al.\ \cite{Rie84};  Bai \cite{Bai03}).
Oliver et al. (1998)
have examined temporally coincident datasets
and discovered  that an identical periodicity is present    
in  the occurrence of sunspots.  

Our research on short-term stellar periodicities
(Massi et al. 1998; Massi et al. 2002) is here
extended to adress  the following question:
 does the periodical radio flaring activity  in \ux  
 imply a periodical spot occurrence?
With this aim we have performed a  timing analysis 
on two 9-year datasets 
of radio and optical observations.  
Section \ref{data} describes the two data bases, 
Sect. \ref{methods} the   timing analysis,
and  finally, 
Sect. \ref{conclusions}  presents our conclusions.

\section{The Data}
\label{data}
The Effelsberg 100-m-telescope observed \ux over a 
frequency range 
1.4--43 GHz
for  almost 3400 days  with 
 quite irregular sampling.
The largest gap in the Effelsberg data base 
is filled by the Green Bank Interferometer (GBI) data base
 at 2.25/8.3 GHz with at least 2 observations per day.
Fig. \ref{fig:fig1}\,d  shows  the  radio observations and Table 1
gives their relative  time schedules.
A visual inspection of Fig. \ref{fig:fig1}\,d   shows   several  large
 flares (S $\ge 250 $mJy).
\begin{table*}
\caption[]{Log of  radio and optical  observations. Start and Stop are given in  Julian Date JD$-$2448965.17 (1992 Dec 8).}
\begin{center}
\begin{tabular}{lcccc}
\hline \hline \noalign{\smallskip}
&Time interval  & Telescope &Data Points& References    \\
\noalign{\smallskip} \hline \noalign{\smallskip}
Radio&0$\div$1000 & Effelsberg 100-m  & 230 &Massi et al. 1998 \\
&1000$\div$3386.95    &Effelsberg 100-m   &211& This work  \\
&1732.71$\div$2858.80 & Green Bank Interferometer  &2533 &Richards et al. 2003 \\
Optical&1.59$\div$3393.46    &Fairborn Automatic Photoelectric Telescope   &869& Aarum Ulv{\aa}s  \& Henry \cite{Aar03}    \\
\noalign{\smallskip} \hline
\end{tabular}
\end{center}
\label{table:log}
\end{table*}
 Massi et al.  (1998)   analyzed the  data
set  (JD 2448965.17+) 0$\div$1000 and determined  
periods of 14, 25, 159 and  56 (circular polarization data) days.
 Richards et al.  (2003) 
analyzing the GBI data 
identified periods of 53, 83, 141 and 280-300 days.
This is very similar to the Sun, where  together with the 
Rieger period  (152-158 days)   several other periodicities 
with time scales of  months appear and disappear  and  
slightly change in  frequency (see references in Oliver \&
Ballester 1995;  Bai 2003).
Here, we  analyze   the full available radio data set, i.e., the 
above mentioned Effelsberg 100-m telescope data 
and the GBI monitoring and further
new  Effelsberg data (see Table 1).  
The  photometric observations of UX Arietis used here
are data taken with the T3 0.4-metre Automatic Photoelectric Telescope
 at Fairborn
Observatory by 
Aarum Ulv{\aa}s  \& Henry (\cite{Aar03}) and consist 
of 869 V band measurements over 9 years 
(see  Fig. \ref{fig:fig1}\,a and Table 1).

\section{Methods of Timing Analysis: Wavelets and  PDM}
\label{methods}
Wavelet analysis decomposes  a one-dimensional time series into a
 two-dimensional time-frequency space and displays the power spectrum 
in a two dimensional color-plot presenting  how the Fourier periods (y) 
vary in time (x) 
(Torrence \& Compo \cite{Tor98}). 
\begin{figure}[htp]
\begin{center}
\includegraphics[angle=0, scale=0.4]{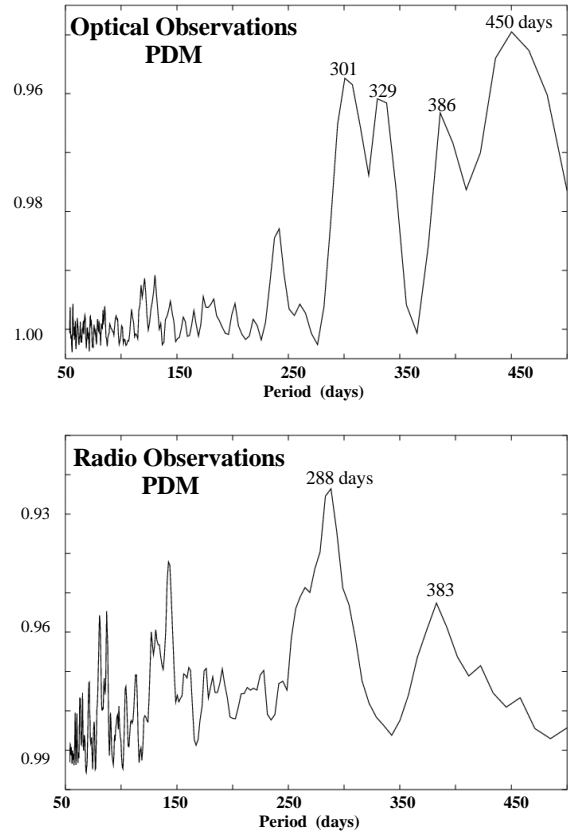}
\caption{Result of the Phase Dispersion Minimization analysis.
Note that the analysis
provides the most probable period as a minimum
 (Stellingwerf \cite{Ste97}) and
the vertical scale is shown  reversed here.
}
\label{fig:fig2}
\end{center}
\end{figure}
The Morlet wavelet results are presented  
in  Fig.~\ref{fig:fig1}\,b for the optical data and
in  Fig.~\ref{fig:fig1}\,c for the radio data.
The power of the frequency component is in arbitrary units
(red stands for the dominant one).
The black contour lines give the 90\%
confidence level obtained by assuming a red
noise background spectrum (Torrence \& Compo \cite{Tor98}).
Dominant periodicities  for both radio and optical data
are evident  in the range 256-512 days
(red area). 
In addition, the radio data show  power at shorter time scales 
(i.e. $< $256 days). 
These are the periodicities 
discussed in Massi et al. (1998) and Richards et al. (2003). 
In the optical data 
these shorter periodicities are significant 
(within black contour lines) but with little power,
therefore, they will not be taken into account in the following discussion
which focuses on the periodicities occurring 
in the interval of 256-512 days. 
\begin{figure}[htp]
\includegraphics[height=0.45\textwidth, width=0.35\textwidth, angle=-90]{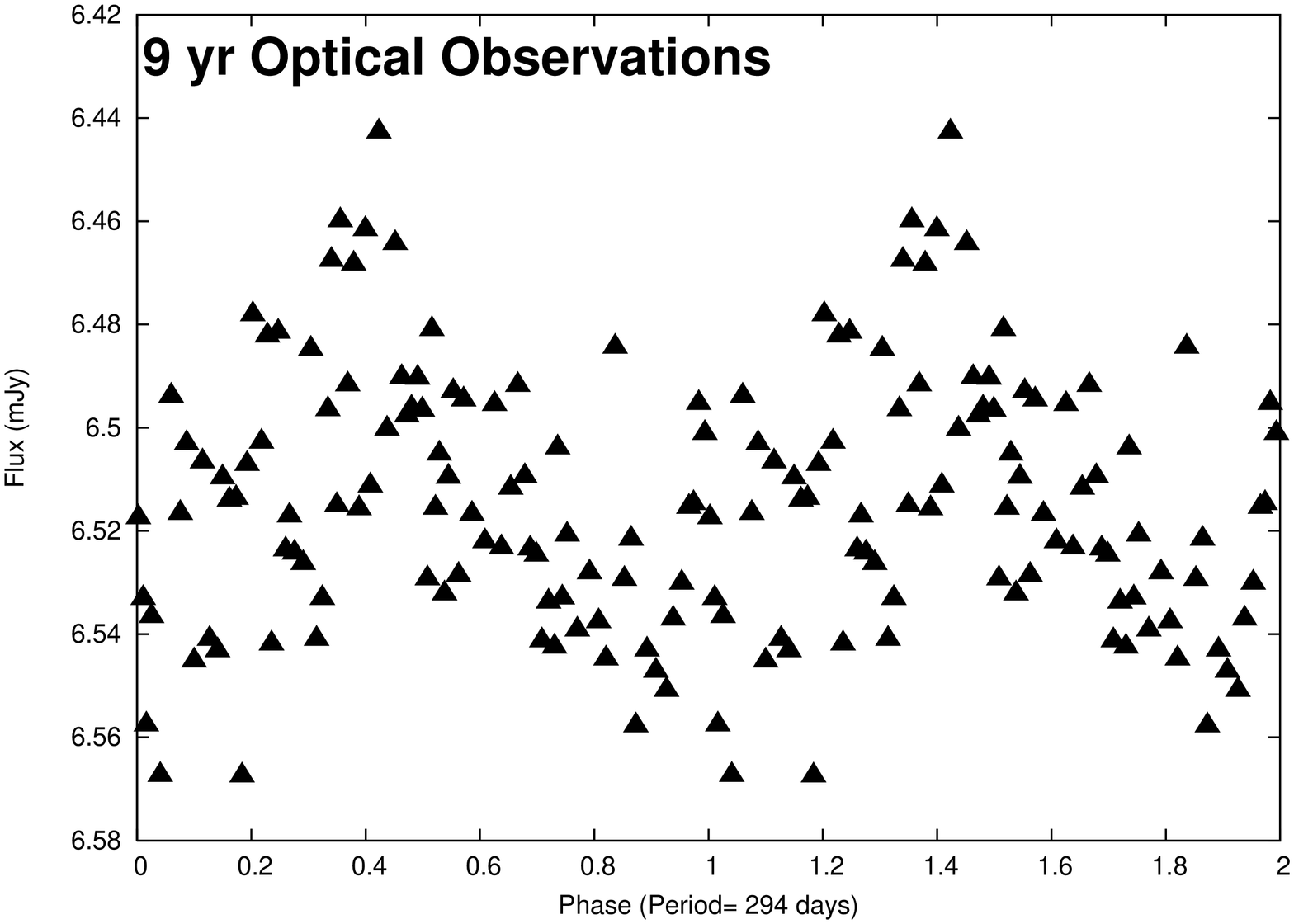}
\includegraphics[height=0.45\textwidth, width=0.35\textwidth, angle=-90]{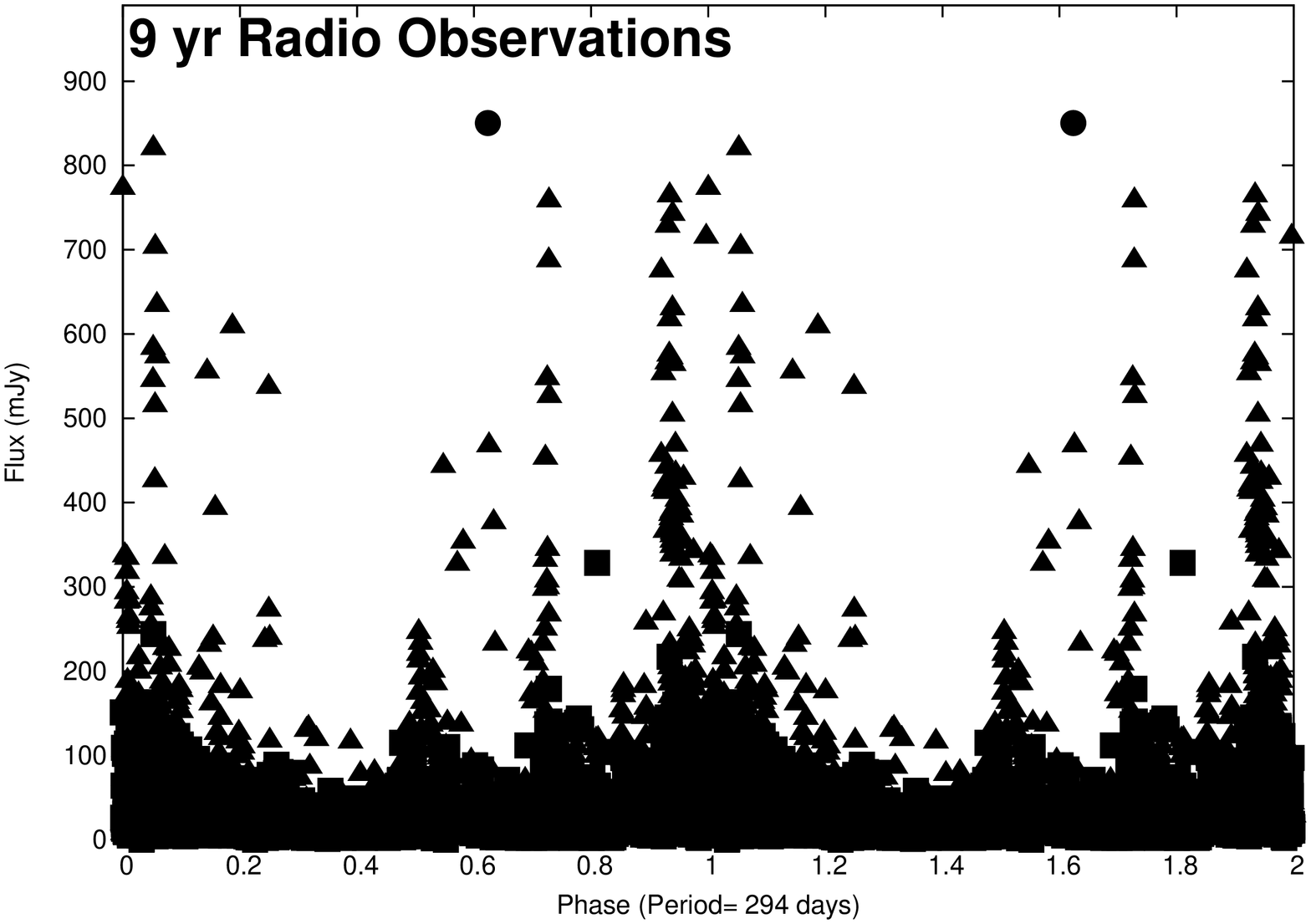}
\caption{
Bottom:
Folding of  the radio  data
with  a period of 294 days.
Phase 0 refers to t$_0$ as in Table \ref{table:log}.
The phase interval 0--1 is repeated twice.
There is a clear lack of energetic flares 
at $\Phi$=0.4.
Top:
Folding
of the photometric V data 
with the same  period of 294~days, as the radio data.
The data are averaged over
phase bins of 0.01.
The peak for V at  $\Phi$=0.4  corresponds to a minimum in spot coverage
and is well synchronized with the ``hole'' of radio flaring 
activity.
\label{fig:fig3}}
\end{figure}


The data have been further analyzed  with the
 Phase Dispersion Minimization method
(PDM, Stellingwerf \cite{Ste97}), which is one of the most 
efficient methods for use with irregularly spaced data.
The  results are plotted in Fig. \ref{fig:fig2}.
The analysis shows that 
the two radio periods, 
 P$_{\rm radio_1}=288\pm6$ days and P$_{\rm radio_2}= 383\pm9$ days,
both have  an optical counterpart within the respective  errors,
i.e., P$_{\rm opt_1}=301\pm7$ days and P$_{\rm opt_2}= 386\pm$11 days.
The influence of noise in the power spectrum and the estimate
of the significance of the determined peaks 
have been established using
the method of Fisher randomization 
(Nemec \& Nemec \cite{Nem85}),  implemented in the Starlink
software package PERIOD. This  resulted in 
false alarm probabilities below 0.01 with 95\% confidence
 for all four periods.
However, due to gaps in the sampling of the optical data,
 we discard P$_2$.  This is because   
 if the optical data are folded with P$_{\rm opt_2}$,   
there is a lack of data over  a consistent  phase 
interval resulting in an artificial  minimum  
(see the regular holes in sampling in
Fig. \ref{fig:fig1}\,a).
This is completely different in the case of the  first period 
of 288-301 days. Figure  \ref{fig:fig3} presents the data folded with
a period of P=294 days, which is
the best common period for both data sets  (radio and optical data).
First, all phase intervals (see Fig. \ref{fig:fig3}) are filled with data
 and no obvious lack of data is present in the light curves.
 Second, optical data (Fig. \ref{fig:fig3} top) show a maximum (i.e.,
a minimum  spot coverage) at
 phase $\Phi$=0.4
exactly  where the radio data
(Fig. \ref{fig:fig3} bottom) show a minimum (i.e. minimum in flaring 
activity).
\section{Discussion and Conclusions}
\label{conclusions}
The results of our timing analysis of a data base of 
radio measurements (Fig \ref{fig:fig1}\,d) and a temporally coincident 
data base of photometric V observations (Fig \ref{fig:fig1}\,a) of
UX Arietis, are:
\begin{enumerate}

\item
The Wavelet analysis over a total time interval of 3400 days
 shows the presence of a quasi-periodic
oscillation  
in the range 256-512 days in both optical and radio data
(Fig \ref{fig:fig1}\,b and  c).

\item
The  PDM analysis of the radio data determines that the  quasi-periodic
oscillation between  256--512 days consists  of two distinct periods: 
P$_{\rm radio_1}=288\pm6$ days (the dominant one) and  
P$_{\rm radio_2}= 383\pm9$ days
(Fig \ref{fig:fig2}).
The PDM analysis of the optical  data  determines
a period  P$_{\rm opt_1}=301\pm7$ days,
coincident within the errors with
the radio periodicity P$_{\rm radio_1}$, 
and also a  period  P$_{\rm opt_2}= 386\pm$11 days,   coincident
with the radio-periodicity  P$_{\rm radio_2}$.
Therefore, both  periods of the radio flaring activity 
have corresponding  optical counterparts.

\item
The analysis of the data folded with the two determined periods
confirms and  corroborates P$_1$ whereas P$_2$ appears to be biased by the gaps in the sampling.
When the data are  folded in phase
with   P=294 days (the best common, optical and radio, period for P$_1$)
 show a good phase coverage (Fig \ref{fig:fig3}).  
 Moreover, there exists a relationship between
the  radio curve and the optical  curve:  
The  maximum V magnitude  (i.e. the minimum spot coverage) is
synchronized  
with the  minimum in  radio flaring activity.
Maximum V magnitude and minimum radio flux density occur both  at
phase  $\Phi=0.4$ (Fig \ref{fig:fig3}).
This close relationship between  two completly independent curves makes
it very likely that the  294-day cycle is real.

\end{enumerate}

On the basis of these results  we conclude
that a periodicity seems to be present on UX Arietis.
This 294-day cycle 
cannot  be an artifact related to a possible spot migration.
In fact, it 
would imply an orbital phase migration rate of 1.24  yr$^{-1}$
whereas a maximum value of 0.26 yr$^{-1}$ has been measured  
(Aarum Ulv{\aa}s  \& Henry \cite{Aar03}).
We are therefore  faced with a real intrinsic stellar periodicity.

As already mentioned in the previous sections,
periodicities with a time scale of months   
are well established 
on the Sun;  they are called 
``Rieger periodicities'' from the name of their discoverer
 (Rieger et al.\ \cite{Rie84}).
They have 
been observed in $\gamma$-ray, X-ray (Rieger et al.\ \cite{Rie84})
and
H$\alpha$ flares (Ichimoto et al.\ \cite{Ich85}),
in flares at radio wavelengths (Bogart \& Bai \cite{Bog85}),
in daily sunspot areas (Oliver et al. 1998)  and  numbers (Ballester et al 1999) and even  
in variations of the solar neutrino flux
(Sturrock et al.\ \cite{Stu99}).

It has been suggested that Rieger periodicities in
solar neutrino data may be due to r-modes (Rossby waves) moving magnetic
regions in and out of the path of neutrinos propagating from the solar
core to
the Earth, assuming an influence of the magnetic field on the
propagation of the neutrinos (Sturrock \cite{Stu03})
possibly by the resonant
spin flavor
precession mechanism
( Akhmedov 1997; Pulido \& Akhmedov \cite{Pul00}).
Moreover, and of interest for the flare production considered here,
equatorially trapped Rossby-type waves
might modulate the emergence of  magnetic flux 
(Lou \cite{Lou00}).
In this respect (Ballester et al. 1999), 
periodic emergence
of magnetic flux through the photosphere may 
operate in two different ways, either forming spots within already
established  active regions (and therefore inducing flares
by magnetic reconnection)
or by forming new spots away from active regions.
Our result (the same period of 294 days in 
temporal  coincidence for  optical  and radio data) is in good agreement
with the first of the two cases, i.e. the  periodical emergence of
magnetic flux occurs in a defined geometrically localized area.
The suggested scenario is therefore as follows:
 During the initial  phase (of 
the 294-day cycle) the spotted surface progressively increases 
because of the emersion of new magnetic structures.
The  emergence area however  remains 
roughly localized (Ballester et al. 1999; Lanza et al. 2001),
so that  new and older magnetic structures interact with each other
and  large (S$\ge$ 250 mJy) 
flares can be observed ($\Phi$ in the range 
  $0.6 \div 1.2$ in Fig. \ref{fig:fig3}-bottom).
Then the minimum of the cycle follows, i.e 
the emersion  of  magnetic flux decreases or stops. Because
there are no new spots and the
the older spots dissipated, 
the optical light curve has its maximum ($\Phi=0.4$ in 
Fig. \ref{fig:fig3} top).
The radio flaring activity dramatically drops
and the radio curve shows a clear ``hole" or lack of flaring
activity, exactly at 
the same phase ($\Phi=0.4$).

\begin{acknowledgements}
We wish to thank Karl M.\  Menten,  Matthias Kadler and Edward  Polehampton
for their  careful reading of the manuscript and
their comments. The results are based on observations with the 100-m
 telescope of the Max-Planck-Institut f\"ur Radioastronomie at Effelsberg.
The Green Bank Interferometer is a facility of the National
Science Foundation operated by the NRAO in support of
NASA High Energy Astrophysics programs.
 This research made use of the data provided by V.  Aarum Ulv{\aa}s
 to the SIMBAD database, operated at CDS, Strasbourg, France.
Wavelet software was provided by C. Torrence and G. P. Compo, and available at URL:
 \verb|http://paos.colorado.edu/research/wavelets/|.
We acknowledge the data analysis facilities provided by the
Starlink Project which is run by CCLRC on behalf of PPARC.
\end{acknowledgements}


\begin{thebibliography}{}

\bibitem[2003]{Aar03}
Aarum Ulv{\aa}s, V.  \& Henry, G. W.
2003, \aap, 402, 1033 

\bibitem[2003]{A3}
 Akhmedov, E. Kh.  
1997, Proc. 4th International Solar Neutrino Conference
(Heidelberg, Germany) hep-ph/9705451

\bibitem[2003]{Bai03}
Bai, T. 
2003,
\apj, 591, 406 

\bibitem[2003]{Bai03}
 Ballester J. L., Oliver, R., \& Baudin, F. 1999, ApJ, 522, L153

\bibitem[2000]{Bea00}
Beasley, A. J. \& G\"udel M.
2000, \apj, 529, 961 

\bibitem[1985]{Bog85}
Bogart, R. S. \&  Bai, T.
1985, \apj, 299, L51 

\bibitem[1985]{Del5}
Budding, E., Lim, J., Slee, O. B.\& White, S. M. 2002,
New Astronomy, 7,1, 35


\bibitem[1985]{D85}
Hall, D. 1976, in Multiple Periodic Variable Stars, IAU Coll. 29, ed. W. Fitch, (Dordrecht: Reidel), 287

\bibitem[1985]{Ich85}
Ichimoto, K., Kubota, J., 
Suzuki, M., Tohmura, I.\&  Kurokawa, H.
1985, Nature, 316, 422 

\bibitem[1985]{D85}
Lanza, A. F., Rodonò, M., Mazzola, L., \& Messina, S. 2001, A\&A, 376, 1011

\bibitem[2000]{Lou00} 
Lou Y., Q.
2000, \apj, 540, 1102 

\bibitem[1998]{Mas98}
Massi, M.,  
Neidh\"ofer, J., Torricelli-Ciamponi, G. \& Chiuderi-Drago, F.
1998, \aap, 332, 194 

\bibitem[2002]{Mas02}
Massi, M., Menten, K. \&  Neidh\"ofer, J.
2002, \aap, 382, 152 

\bibitem[1985]{Nem85}
Nemec, A. F. \& Nemec, J. M.
1985, \aj, 90, 2317 

\bibitem[1997]{Nis97} 
Nishio, M., Yaji, K., Kosugi, T., Nakajima, H. 
\& Sakurai, T.
1997, \apj, 489, 976 

\bibitem[1998]{Oli98}
Oliver, R., \& Ballester, J. L. 1995, Sol. Phys., 156, 145

\bibitem[1998]{Oli98}
Oliver, R., Ballester, J. L. \& Baudin, F.
1998, Nature, 394, 552 


\bibitem[1985]{Dl85}
Owen, F. N., Jones, T. W. \& Gibson, D. M. 1976, ApJ, 210, L27


\bibitem[2000]{Pul00}
Pulido, J. \&  Akhmedov, E. Kh.
2000, Astroparticle Phys., 13,  227 
%
%


\bibitem[2003]{Ric03}
Richards, M. T., Waltman, E. B., Ghigo, F. D.  \& Richards, D. St. P.
2003, \apjs, 147, 337 

\bibitem[1984]{Rie84}
Rieger, E.  Kanbach, G., Reppin, C., 
et al.
1984,
Nature, 312, 623 

\bibitem[1997]{Ste97}
Stellingwerf, R. F.
1997, \apj, 486, 886 

\bibitem[1999]{Stu99}
Sturrock, P. A., Scargle, J. D., Walther, G. \&  Wheatland, M. S.
1999, \apj, 523, L177 

\bibitem[2003]{Stu03}
Sturrock, P. A.
2003, Am.\ Astron.\ Soc.\ Solar Physics Division Meeting 34.08.05

\bibitem[1998]{Tor98}
Torrence, C. \& Compo, G. P.
1998, Bull.\ Am.\ Meteo.\ Soc., 79, 61 
\end{thebibliography}
\end{document}